\documentclass[aps,prl,nofootinbib,twocolumn,showpacs,floatfix,superscriptaddress]{revtex4}

\usepackage{amsfonts,amssymb,epsfig,verbatim,mathbbol,amstext,amsmath,psfrag}

\def\eps{\epsilon}
\def\pa{\partial}
\def\Im{{\rm Im}\,}

\begin{document}

\title{Instanton constituents in the O(3) model at finite temperature}

\author{Falk Bruckmann}
\affiliation{Institut f\"ur Theoretische Physik, Universit\"at Regensburg, D-93040 Regensburg, Germany}

\begin{abstract}

It is shown that instantons in the $O(3)$ model at finite temperature 
consist of fractional charge constituents 
and the (topological) properties of the latter are discussed.

\end{abstract}

\pacs{11.15.Ha}
\keywords{keywords appear here}

\maketitle


\section{Introduction}

The famous $O(3)$ model in two dimensions has various applications in condensed matter 
and high energy physics. 
From the point of view of gauge field theories, 
it is one of the few toy models that exhibits asymptotic freedom 
and a dynamical mass generation.

The other main feature of the $O(3)$ model is the existence of {\em solitonic solutions} \cite{polyakov:75}. 
They are stabilized by a topological quantum number 
belonging to the second homotopy group of the color two-sphere\footnote{
because every configuration of finite action 
can be 
compactified by including spatial infinity}. 

The solitons of any topological charge are known explicitly 
making use of the {\em complex structure} of the Bogomolnyi equation, see below.
All these properties, including the new findings in this letter, find immediate generalizations in $CP(N)$ models.

That the soliton of unit topological charge can be parametrized by two locations (see Eq.~(\ref{eqn_inst_rat_para})),
has initiated speculations, whether it is actually made of two constituents, 
named `instanton quarks' \cite{belavin:79}. 
In the profile of the topological charge (or action) density, however, there is no sign of the constituents, 
they rather generate one lump. 
On the other hand, the measure on the moduli space of classical solutions 
can be written in terms of the constituent locations \cite{fateev:79_berg:79_diakonov:99}.

At this point I invoke some knowledge from gauge theories in four dimensions: 
Yang-Mills instantons at finite temperature, called calorons \cite{harrington:78},
have magnetic monopoles as constituents,
provided the holonomy is nontrivial \cite{kraan:98a_lee:98b}. 
These calorons are obtained by two effects. 
One is to squeeze instantons by compactifying the time-like direction 
to the usual circle of circumference $\beta=1/k_B T$. 
The second ingredient is the different color orientation of the instanton copies along that compact direction 
(for a recent review see \cite{bruckmann:07a}).
A similar picture applies to
doubly periodic Yang-Mills instantons \cite{ford:02a}. 

The $O(3)$ model has been investigated on the two-torus, 
too \cite{richard:83_aguado:01}.
On $R^2$, fractional charge objects only exist at the expense of singularities \cite{gross:78_zhitnitsky:89}. 

In this Letter I will show that large instantons in the $O(3)$ model at finite temperature
{\em dissociate into two constituents},
provided one allows for a nontrivial transition function in the time-like direction 
being part of the global symmetry of the model. 
The constituents are static and of (in general different) {\em fractional topological charge}
governed by the new holonomy parameter.

\section{
The model and its solitons on $R^2$}

Conventionally, the $O(3)$ model is defined in terms of a
three-vector $\phi^a$,
taking values on a two-sphere $S^2_c$ in color space, 
$\phi^a\phi^a=1,\, a=1,2,3$ (the sum convention is used).
Its action is just the usual kinetic term, whereas the 
integer-valued topological charge reads
\begin{equation}
Q=\frac{1}{8\pi}\int d^2 x\, \eps_{\mu\nu}\eps_{abc}\phi^a\pa_\mu\phi^b\pa_\nu\phi^c
\in\mathbb{Z}\,,\quad
\mu=1,2\,.
\label{eqn_inst_Qphi}
\end{equation} 
It is useful to introduce a complex structure,
\begin{equation}
\frac{\phi^1+i\phi^2}{1-\phi^3}=u(z,z^*)\,,\quad z=x_1+i x_2\,.
\label{eqn_inst_wintro}
\end{equation}
The zeroes and poles of $u$ have the immediate interpretation of $\phi$ being on the north and south pole of $S^2_c$,
respectively. 
Such a behavior is necessary for the field to have a winding number, since it has to fully cover $S^2_c$.

Configurations of minimal action in $S\geq 4\pi|Q|$ fulfil first order `selfduality' equations 
{\em solved by $u$ being a function of $z$ (or $z^*$) alone}.
The topological charge in terms of $u(z)$ reads
\begin{equation}
Q=\int d^2x\, q(x)\,,\quad q(x)=\frac{1}{\pi}\frac{1}{(1+|u|^2)^2}\left|\frac{\pa u}{\pa z}\right|^2\,.
\label{eqn_inst_Qu}
\end{equation}

Concerning the
solutions, the meromorphic ansatz,
\begin{equation}
u(z)=\frac{\lambda}{z-z_0}\,,
\label{eqn_inst_wmero}
\end{equation}
has a pole at $z_0$ and a zero at infinity, which makes it plausible that this configuration has charge 1.
Indeed, the profile of the topological charge density
\begin{equation}
q(x)=\frac{1}{\pi}\frac{\lambda^2}{(|z-z_0|^2+\lambda^2)^2}
\label{eqn_inst_qprofile}
\end{equation} 
integrates to $Q=1$.
One recognizes the size $\lambda$ and the location $z_0$ of the instanton.

The model has a global $O(3)$ symmetry rotating $\phi^a$ 
(the topological charge $Q$ from Eq.~(\ref{eqn_inst_Qphi})
for example is a triple product and thus a pseudoscalar under this symmetry).
An $SO(2)$ subgroup of rotations of $(\phi^1,\phi^2)$ 
acts on $u$ by multiplication with a complex phase,
which leaves $q$ unchanged, see Eqs.~(\ref{eqn_inst_wintro}) and (\ref{eqn_inst_Qu}).

An analytic ansatz $1/u$ with the roles of pole and zero 
(i.e. north and south pole on $S^2_c$) interchanged 
gives the same profile $q(x)$.
The function $u(z)$ can also have both its zero and pole at finite $z$, 
e.g.\ in the rational function
\begin{equation}
u_{\rm rat}(z)=\frac{z-\hat{z}}{z-\check{z}}\,.
\label{eqn_inst_rat_para}
\end{equation}
This reparametrization offers the possibility of constituents at locations $z=\{\hat{z},\check{z}\}$.
However, the topological density still has one lump; 
it is of the form (\ref{eqn_inst_qprofile}) 
with size $\lambda=|\hat{z}-\check{z}|/2$ around the center of mass $z_0=(\hat{z}+\check{z})/2$.

\section{The case of finite temperature}

Higher charge solutions are given by a simple product ansatz
$\prod_{k=1}^Q\lambda/(z-z_{0,k})$.
For a solution in the finite temperature setting, i.e.~identifying $z\sim z+i\beta$, 
one would have to consider the infinite product
\begin{equation}
\prod_{k=-\infty}^{\infty}\frac{\lambda}{z-z_0-ik\beta}\,,
\label{eqn_prod_inf}
\end{equation}
which is infinite.
Neglecting an infinite factor and
using the product representation of the sine function, 
a regularized version of (\ref{eqn_prod_inf}) is
\begin{equation}
\frac{\lambda}{\sinh((z-z_0)\frac{\pi}{\beta})}\,.
\label{eqn_cal_naivereg}
\end{equation}
This function  
changes sign 
under $z \to z+i\beta$.

More systematically, let $u(z)$ be a function with simple poles at $z=z_0+i k\beta$. 
Then the Mittag-Leffler theorem fixes the singular part of $u(z)$ uniquely, 
provided the residues at these points are given.
For a periodic $u(z)$ all the residues must be the same, which gives
$
\lambda/(\exp((z-z_0)\frac{2\pi}{\beta})-1)
$
up to an analytic part.

\begin{figure*}
\includegraphics[width=0.32\linewidth]{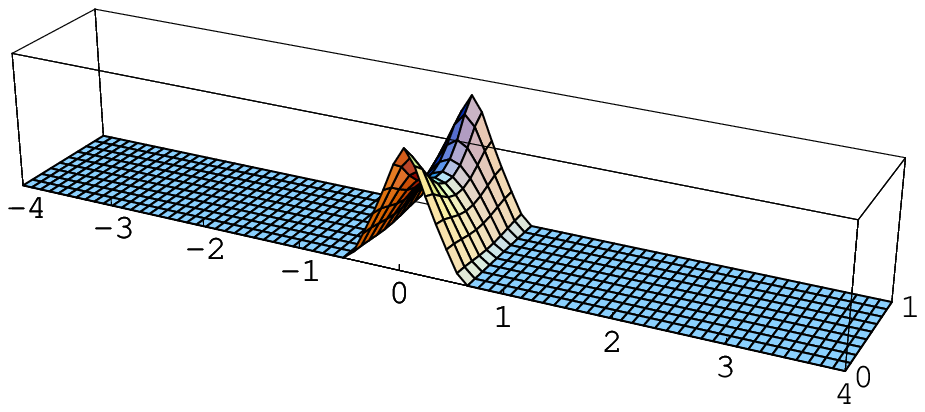}
\includegraphics[width=0.32\linewidth]{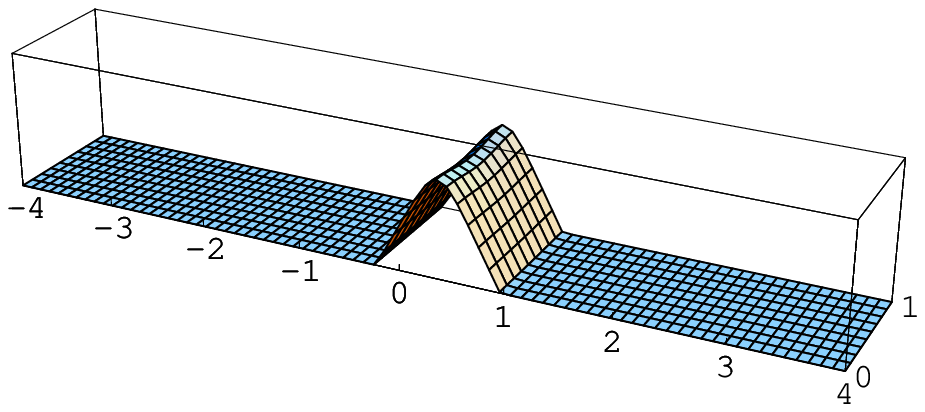}
\includegraphics[width=0.32\linewidth]{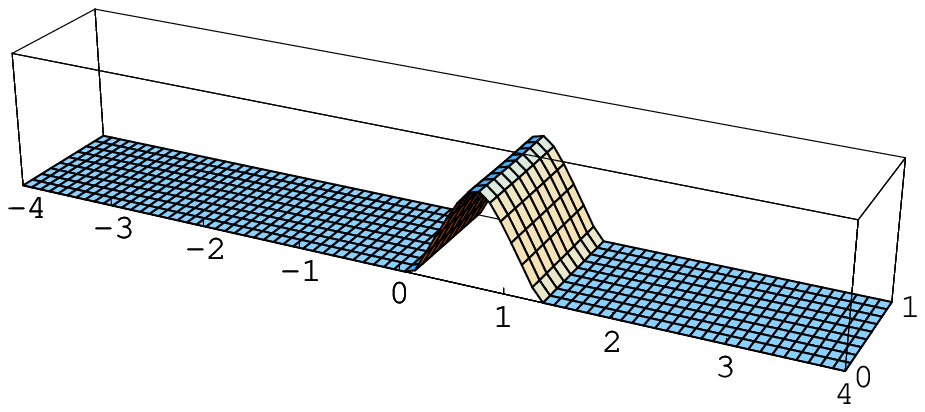}\\
\includegraphics[width=0.32\linewidth]{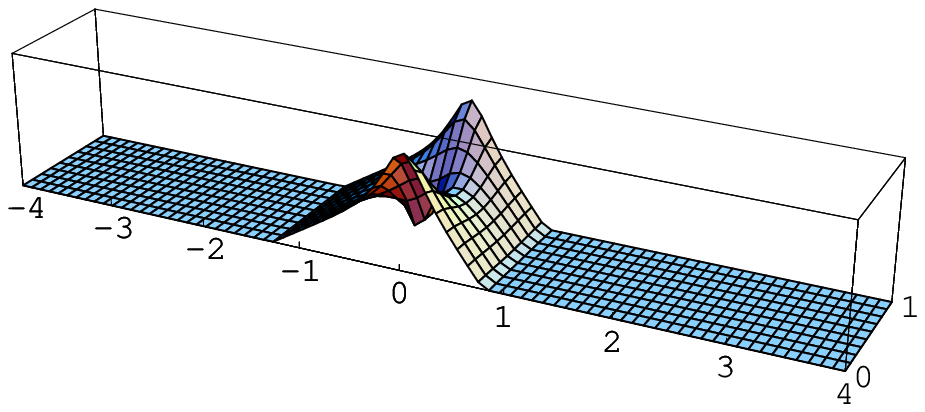}
\includegraphics[width=0.32\linewidth]{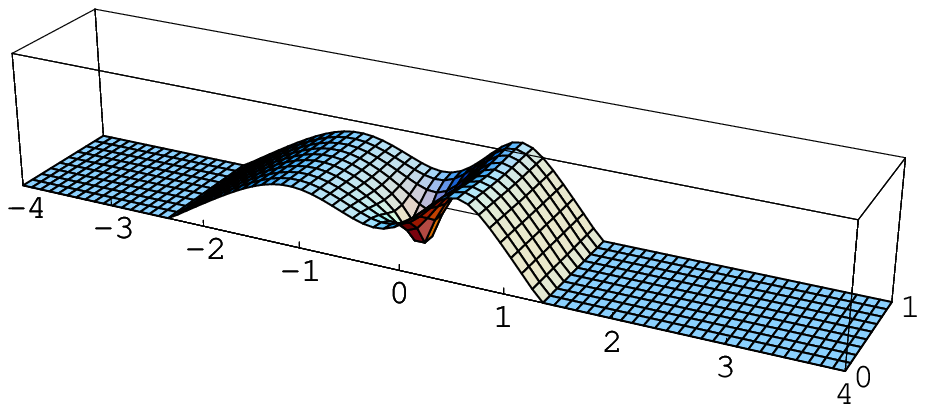}
\includegraphics[width=0.32\linewidth]{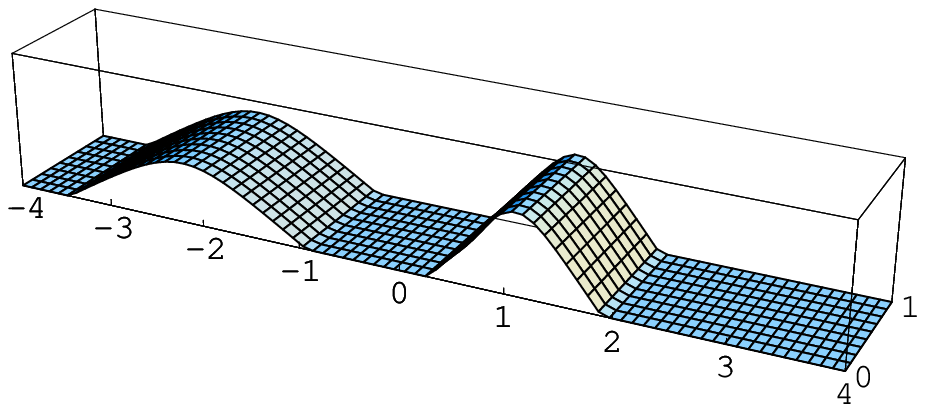}\\
\includegraphics[width=0.32\linewidth]{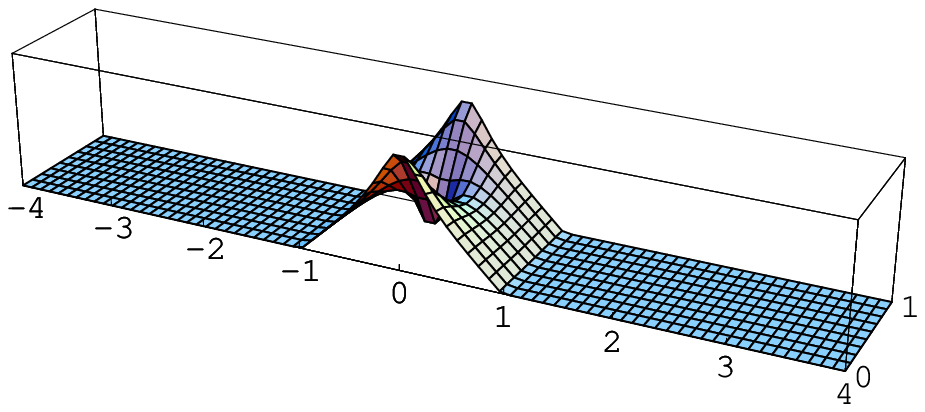}
\includegraphics[width=0.32\linewidth]{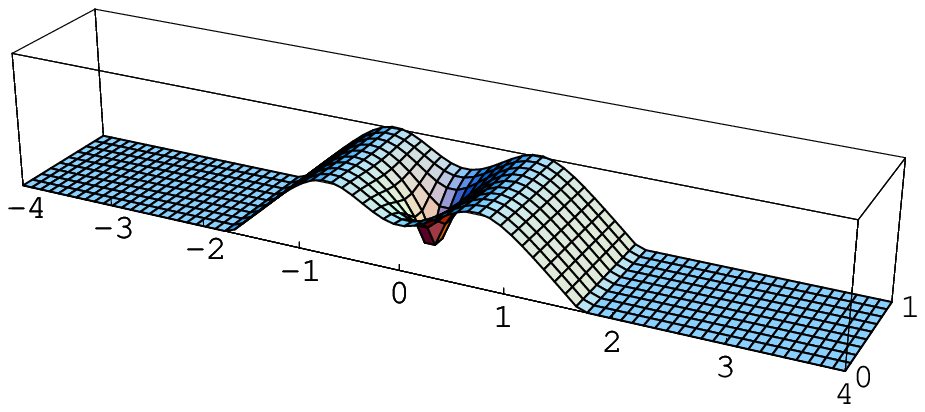}
\includegraphics[width=0.32\linewidth]{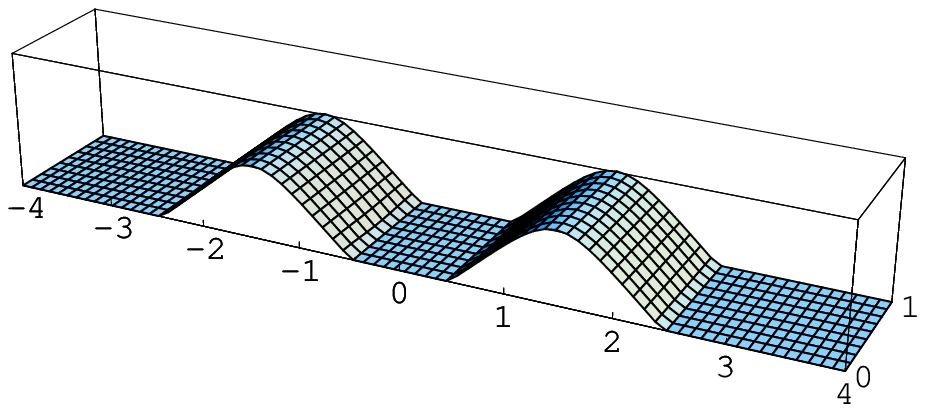}
\caption{Logarithm of the topological density of solitons 
with different size and holonomy parameter
(plugging (\ref{eqn_cal_u}) into (\ref{eqn_inst_Qu}) and cut off below $e^{-5}$).
From left to right $\lambda=1,\,10,\,100$ (with locations growing like $\ln \lambda$ 
according to Eq.~(\protect\ref{eqn_const_locs})).
From top to bottom the periodic case $\omega=0$ with one lump
(the massless constituent being infinitely spread 
like for the Harrington-Shepard caloron),
an intermediate case, $\omega=1/3$, 
and the antiperiodic case $\omega=1/2$ with identical constituents.}
\label{fig_densities}
\end{figure*}

Utilizing the $SO(2)$ symmetry, let $u(z)$ be {\em periodic up to a phase} $\exp(2\pi i\omega),\: \omega\in [0,1]$.
This phase fixes the relative residues at consecutive poles and the new solution is
\begin{equation}
u(z;\omega)=\frac{\lambda\exp(\omega(z-z_0)\frac{2\pi}{\beta})}{\exp((z-z_0)\frac{2\pi}{\beta})-1}\,.
\label{eqn_cal_u}
\end{equation} 
It is this solution that {\em gives rise to instanton constituents} for large $\lambda$.

This phenomenon is in agreement with the large size regime of Yang-Mills calorons. 
The parameter $\omega$ gives the complex orientation of the residues in $u(z;\omega)$ along $\Im z$,
corresponding to the color orientations of Yang-Mills instanton copies in the ADHM formalism, 
and will therefore be called holonomy parameter.
The only difference to the case of nonabelian gauge theories is that 
there the nonperiodicity of the primary object,
the gauge field $A_\mu(x)$,  can be compensated by a time-dependent gauge transformation, 
whereas in the $O(3)$ model this cannot be done as the gauge symmetry is global.

\section{Constituent properties}

Vanishing parameter $\omega$ 
(as well as the equivalent $\omega=1$) refer to the periodic case, 
while $\omega=1/2$ is the antiperiodic case 
with $u(z;1/2)$ agreeing with Eq.~(\ref{eqn_cal_naivereg}) up to a factor 2. 
As Fig.~\ref{fig_densities} shows,
the periodic case consists of one lump of action density, 
while the antiperiodic soliton dissociates into two identical lumps, 
when the size parameter $\lambda$ is large.
Moreover, these constituent lumps are almost static, i.e.\ Im $z$-independent 
(although $u(z)$ is not).

The general case of holonomy parameter $\omega$ reveals lumps of `masses' 
(that is energy density integrated along Re $z$) of $\omega/\beta$ 
and $\bar{\omega}/\beta,\:\bar{\omega}\equiv 1-\omega\in[0,1]$,
in complete analogy to the YM caloron constituents. 
Their sum multiplied by the Im $z$-extension $\beta$ gives $Q=1$.

This can be best understood by rewriting Eq. (\ref{eqn_cal_u}) into
\begin{equation}
u(z;\omega)=\frac{1}{\exp(\bar{\omega}(z-z_2)\frac{2\pi}{\beta})-
\exp(-\omega(z-z_1)\frac{2\pi}{\beta})}
\label{eqn_cal_u_rewritten}
\end{equation}
with locations 
\begin{equation}
z_1=z_0-\beta\,\frac{\ln\lambda}{2\pi\omega}\,,\quad
z_2=z_0+\beta\,\frac{\ln\lambda}{2\pi\bar{\omega}}\,.
\label{eqn_const_locs}
\end{equation}
For $\lambda>1$ the ordering is Re $z_1<$ Re $z_0<$ Re $z_2$.

That these locations are really the centers of constituents can be seen for large sizes $\lambda\gg 1$, 
where Re $z_2\gg$ Re $z_1$, such that around $z_{1,2}$ 
one of the terms in Eq.~(\ref{eqn_cal_u_rewritten}) is exponentially suppressed leading to
\begin{equation}
\lambda\gg 1:\:\: u(z;\omega)\simeq\left\{\begin{array}{ll}
\exp(-\bar{\omega}(z-z_2)\frac{2\pi}{\beta}) & \mbox{for } z\simeq z_2\,,\\
-\exp(\omega(z-z_1)\frac{2\pi}{\beta}) & \mbox{for } z\simeq z_1\,.
\end{array}\right.
\end{equation} 
The corresponding constituent solution 
\begin{equation}
u_{\rm const}(z;\omega)=\exp(\omega z \, \frac{2\pi}{\beta})\,,
\label{eqn_const_1}
\end{equation} 
gives rise to an exponentially localized profile 
\begin{equation}
q(x)=\frac{\pi\omega^2}{\beta^2 \cosh^2(\omega \mbox{ Re\,} z\,\frac{2\pi}{\beta})}\,,
 \end{equation}
which is static and yields
$Q=\omega$. 
The other constituent 
$
\exp(-\bar{\omega} z \, \frac{2\pi}{\beta})
$
has the same profile with $\omega$ replaced by $\bar{\omega}$. 
Actually, $u_{\rm const}(z,\omega+m)$ with any integer $m$ is consistent with the boundary condition 
and gives
$Q=|\omega+m|$.

In the far field limit $|$Re $z|\to\infty$ again one of the terms 
in Eq.~(\ref{eqn_cal_u_rewritten}) is exponentially small and by neglecting it,
only the nearest constituent is visible
(for Re $z\to\mp\infty$ the one at $z_{1,2}$), 
in contrast to the YM case, where all constituents are present via algebraic tails.

\begin{figure}[!h]
\psfrag{i}{Re $z\to\infty$}
\psfrag{m}{Re $z\to-\infty$}
\psfrag{b}{Im $z=\beta\Rightarrow$}
\psfrag{c}{arg $u(z)=2\pi\omega$}
\psfrag{z}{Im $z=0$}
\psfrag{r}{Re $z=$ const.}
\includegraphics[width=0.7\linewidth]{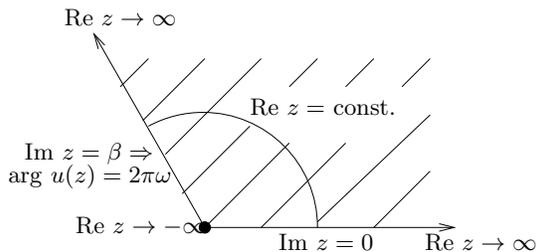}
\caption{The image of the function $u_{\rm const}(z;\omega)$, Eq.~(\ref{eqn_const_1}),
of a single constituent covers a fraction $\omega$ of the complex plane.}
\label{fig_fract_cover}
\end{figure}

The fractional charge finds its counterpart 
in the fractional covering of the complex plane by 
$u(z;\omega)$, see Fig~\ref{fig_fract_cover}.

\section{Topology}

For the topological description of the possible solitonic solutions
I start with the usual argument
that a finite topological charge demands an Im $z$-independent function $u(z)$ 
asymptotically, i.e.\ for large $|$Re $z|$.
This is consistent with the nontrivial boundary condition, 
the phase change $\exp(2\pi i\omega)$ in Im $z$, 
only if the asymptotic values of $u(z)$ are zero or infinity,
i.e.\ the poles $\phi_3=\pm1$ on the complex sphere\footnote{
These poles are distinguished by the choice of the $SO(2)$ subgroup in the boundary condition.}.

The topological charge density of Eq.~(\ref{eqn_inst_Qphi}),
the (pullback of the) volume form, 
can locally be written as a curl (exterior derivative of a one-form), 
\begin{eqnarray}
q(x)&=&\frac{1}{4\pi} \eps_{\mu\nu}\sin\theta\,\pa_\mu\theta\pa_\nu\varphi
\label{eqn_Q_spher}
=\frac{1}{4\pi} \eps_{\mu\nu}\pa_\mu[(\pm1-\cos\theta)\pa_\nu\varphi]\nonumber\\
&&\phi_3=\cos\theta\,,\:\:\phi_1+i\phi_2=\sin\theta e^{i\varphi}\,,
\label{eqn_Q_curl}
\end{eqnarray}
where the expression in the square bracket is a regular representation 
in spherical coordinates in $\phi$
around the north and south pole, respectively.
Consequently, I now divide the coordinate space into regions,
where $\phi(x)$ is on the northern and southern hemisphere, called $N$ and $S$, respectively, 
and use the corresponding regular expression there. 

Then $Q$ reduces to boundary integrals, both on the preimage of the equator 
separating the hemispheres (there $\cos\theta=0$ or equivalently $|u(z)|=1$)
and on the boundary of space itself, 
\begin{eqnarray}
Q&=&\frac{1}{4\pi}\int_{\phi_3(x)=0} d\vec{\sigma}\left((+1)-(-1)\right)\vec{\pa}\varphi+
\label{eqn_Q_eq}\\
&&\frac{1}{4\pi}\int dx_1 \left.(\pm1-\cos\theta)\frac{\pa\varphi}{\pa x_1} 
\right|^{x_2=\beta}_{x_2=0}\,.
\end{eqnarray}
The second term vanishes since $\varphi(x_1,x_2+\beta)=\varphi(x_1,x_2)+2\pi\omega$ 
and $\theta(x_1,x_2+\beta)=\theta(x_1,x_2)$ (and because the assignment to the hemispheres is the same on both boundaries).
In the first term the curves are the borders of two regions 
with opposite sign in Eq.~(\ref{eqn_Q_curl}), 
endowing the curve with opposite orientations.
Thus, {\em the topological charge $Q$ is given as the sum of oriented changes in the angular variable 
$\varphi=\arctan(\phi_2/\phi_1)$ 
along the `equator lines' $\phi_3=0$ divided by $2\pi$}.

For these equator lines three types of configurations are possible.
The first one is a curve stretching from one boundary of space 
to (the same $x_1$ at) the other boundary. 
It separates $S$ on its lhs.\ from $N$ on its rhs.\ or vice versa.
Such a line picks up a $\varphi$-change of $\pm 2\pi\omega$ up to multiples of $2\pi$. 
Hence, these equator lines contribute fractional topological charges $\omega+m$ or $-\omega+n$. 
For the definiteness of $Q$ 
the integers are restricted, $m\geq 0,n\geq 1$.

\begin{figure}[b]
\psfrag{o}{$\omega$}
\psfrag{p}{$1-\omega$}
\psfrag{z1}{$z_1$}
\psfrag{z2}{$z_2$}
\includegraphics[width=0.8\linewidth]{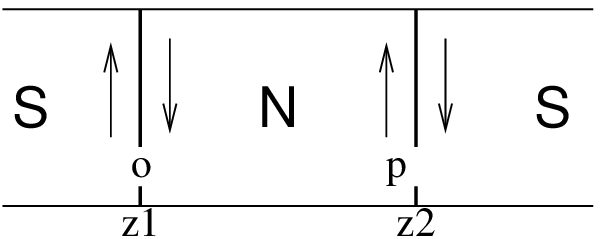}

\vspace{0.5cm}
\psfrag{1}{$\,1$}
\includegraphics[width=0.8\linewidth]{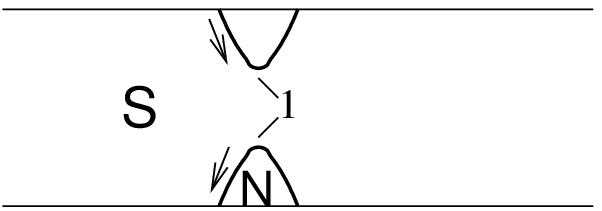}
\caption{The distribution of the southern and northern hemisphere for a large (top) and small (bottom) instanton 
and the contributions to the topological charge $Q$ from the oriented equator lines.
To get the picture for the individual constituents, one simply cuts the upper plot vertically in the middle.}
\label{fig_topology}
\end{figure}

In fact, individual or well-separated constituents are examples of this phenomenon, 
because the equator lines run through their centers. 
Fig.~\ref{fig_topology} top depicts the situation for a large instanton,
where 
$m$ and $n$ take 
values 0 and 1, respectively
(cf. the discussion below Eq.~(\ref{eqn_const_1})).

The second type of equator line returns to the same boundary. 
Because of the periodicity of $\phi_3$, 
there will be another equator line starting and ending at the same $x_1$ at the other boundary.
It is easy to see that the sum of $\varphi$-changes on these two lines is an integer.
A small instanton provides an example of this type of equator line with contribution 1, 
see Fig.~\ref{fig_topology} bottom. 

Finally, a closed curve encircling $N$ or $S$ is the third possibility for an equator line. 
It does not feel the boundary condition and therefore is well-known, 
e.g.\ from the Wu-Yang construction of the Dirac monopole.
The contribution of such an equator line is again an integer, the winding number of $\varphi$
(around a zero or pole of $u(z)$).

The essence of the topological considerations is that 
whenever the preimage of one hemisphere is an `island' in the preimage of the other hemisphere,
there is an integer contribution to the topological charge
(the last two types).
{\em The fractional part of $Q$ emerges from equator lines between the boundaries
separating different hemispheres} 
(for like hemispheres the contributions in Eq.~(\ref{eqn_Q_eq}) have the same sign and cancel).

This has the interesting consequence, that
the fractional charge cannot accumulate -- to say $\pm2\omega$ --
without being accompanied by a contribution $\mp\omega$ inbetween
(including integers $m$ and $n$)
and that the fraction of the topological charge $Q$ is determined purely by the asymptotics $|x_1|\to\infty$.
Different poles there give $\omega$ or $-\omega$ 
(examples are the individual constituents discussed in the previous section),
whereas same asymptotic poles give no fractional charge (like for the instantons, see Fig.~\ref{fig_topology}).

Of course, these calculations cannot determine the local distribution of the topological charge density $q(x)$;
in the examples, however, the latter turns out to be concentrated around the equator lines.

\section{Conclusions}

At finite temperature, 
instantons in the $O(3)$ model reveal fractional charge constituents.
These novel solutions have been given by simple analytic expression using complex functions.
In general, the topological charge consists of a fractional part $\pm\omega$,
where $\omega$ is the holonomy parameter in the boundary conditions governing the masses of the static constituents,
plus possible integers from undissociated instantons being 
 time-dependent.

The size parameter of large instantons transmutes into the distance of its constituents,
which themselves have a size proportional to $\beta$.
This also resolves the puzzle, why 
the instanton quark locations --
even if they are locations of constituents in the zero temperature limit
(which can be achieved) --
do not show up as individual topological lumps:
for $\beta\to\infty$ the constituents become large and inevitably overlap.
Put differently, in the zero temperature case, 
there is no other scale besides the distance of the instanton quarks,
that could localize the latter.

The finite temperature case might be viewed as the limit of elongated space-time tori 
(with nontrivial boundary conditions), 
where the smaller extension governs the size of the constituents, 
which can separate in the direction of the larger extension.

Many of these features are similar to those of Yang-Mills calorons.
It would be interesting to investigate how far this analogy goes, 
in particular since $CP(N)$ models can be parametrized by virtue of a gauge field.
From the corresponding Yang-Mills phenomenon, 
{\em fermions} are 
expected to localize to the constituents in the  background 
and to hop with their boundary conditions.

The constituent picture should also be of relevance for
understanding the physical mechanisms 
of the $O(3)$ model at finite temperature.\\

\noindent The author thanks P.~van Baal and A.~Wipf for discussions and acknowledges support by DFG (BR 2872/4-1).


\end{document}